\title{Local Kernel Dimension Reduction in Approximate Bayesian Computation}
\author{
        Jin Zhou \and Kenji Fukumizu
}
\date{\today}

\documentclass[12pt]{article}
\usepackage{amsmath}
\usepackage{setspace}
\usepackage{bm}
\usepackage{dsfont}
\usepackage[lined,boxed]{algorithm2e}
\usepackage[cal=cm,
calscaled=.96,
bb=ams,
frak=mma,
scr=rsfs]{mathalfa}
\newcommand\independent{\protect\mathpalette{\protect\independenT}{\perp}}
\def\independenT#1#2{\mathrel{\rlap{$#1#2$}\mkern2mu{#1#2}}}
\begin{document}
\maketitle

\begin{abstract}
Approximate Bayesian Computation (ABC) is a popular sampling method in applications involving intractable likelihood functions. Without evaluating the likelihood function, ABC approximates the posterior distribution by the set of accepted samples which are simulated with parameters drown from the prior distribution, where acceptance is determined by distance between the summary statistics of the sample and the observation. The sufficiency and dimensionality of the summary statistics play a central role in the application of ABC. This paper proposes Local Gradient Kernel Dimension Reduction (LGKDR) to construct low dimensional summary statistics for ABC. The proposed method identifies a sufficient subspace of the original summary statistics by implicitly considers all non-linear transforms therein, and a weighting kernel is used for the concentration of the projections. No strong assumptions are made on the marginal distributions nor the regression model, permitting usage in a wide range of applications. Experiments are done with both simple rejection ABC and sequential Monte Carlo ABC methods. Results are reported as competitive in the former and substantially better in the latter cases in which Monte Carlo errors are compressed as much as possible. 
\end{abstract}

\section{Introduction}

Monte Carlo methods are popular tools in sampling and inference problems. While the Markov Chain Monte Carlo methods find
successes in applications where likelihood functions are known up to a unknown constant, MCMC can not be used in scenarios where likelihoods are intractable. For these cases, if the problem can be characterized by a generative model, Approximate Bayesian Computation (ABC) is often a candidate approach. ABC is a Monte Carlo method that approximates the posterior distribution by jointly generating simulated data and parameters and does the sampling based on the distance between the simulated data and the observation, without evaluating the likelihoods. ABC was first introduced in population genetics \cite{Pritchard1999} \cite{Beaumont2002} and then been introduced to a range of complex applications including dynamical systems \cite{Toni2009}, ecology \cite{Csillery2010}, Gibbs random fields \cite{Grelaud2009} and demography \cite{Bertorelle2010}. 

The accuracy of ABC posterior depends on the Monte Carlo errors induced in the samplings. Given the generative model $p(y|\theta)$ of observation $y_{obs}$ with parameter $\theta$, consider summary statistics $s_{obs}=G_s(y_{obs})$ and $s=G_s(y)$, where $G_s: Y\rightarrow S$ is the mapping from the original sample space $Y$ to low dimensional summary statistics $S$. The posterior distribution, $p(\theta|y_{obs})$, is approximated by $p(\theta|s_{obs})$, which is constructed as $p(\theta|y_{obs}) = \int p_{ABC}(\theta,s|s_{obs})ds$, with

\begin{equation} \label{eq:1}
p_{ABC}(\theta,s|s_{obs}) \propto p(\theta) p(s|\theta)K(\| s-s_{obs}\|/\epsilon),
\end{equation}
where $K$ is a smoothing kernel with bandwidth $\epsilon$. In the case of simple rejection ABC, $\it{K}$ is often chosen as an indicator function $I(\| s-s_{obs}\| < \epsilon)$. If the summary statistics $s$ are sufficient, it can be shown that \eqref{eq:1} would converges to the posterior $p(\theta|s_{obs})$ as $\epsilon$ goes to zero\cite{Blum2010}.

As can be seen above, sampling is based on the distance between the summary statistics of the simulated sample $s$ and the observation $s_{obs}$. Approximation errors are induced by the distance measure and proportional to the distance threshold $\epsilon$. It is desirable to set $\epsilon$ as small as possible, but a small threshold will increase the simulation time. This is a trade-off between the accuracy and the efficiency (simulation time) determined by the choices of thresholds and summary statistics. According to recent results on asymptotic properties of ABC \cite{Frazier2016} \cite{Li2016}, assuming that the summary statistics follow the central limit theorem, the convergence rate of ABC when accepted sample size $N \rightarrow \infty$ is depended on the behavior of $\mu=\epsilon d_N$, where $\epsilon$ is the threshold above and the $d_N$ is defined as of same magnitude of $eigen(\Sigma_N)$, the eigenvalue of the covariance matrix of the summary statistics as the function of $N$. In practice, if a specific sampling method is chosen, the threshold $\epsilon$ is constrained by the computing resources and time, thus can be accordingly determined. The design of summary statistics then remains the most versatile and difficult part in developing an efficient ABC algorithm. To avoid the ``curse of dimensionality", summary statistics should be also chosen to be low dimensional in addition of sufficiency. 

A vast body of literature of ABC have been published. Many are devoted to reduce the sampling error by using more advanced sampling method, from simple Rejection method\cite{Moore1995}, Markov Chain Monte Carlo(MCMC)\cite{Marjoram2003} to more sophisticated methods like sequential Monte Carlo \cite{Sisson2007}\cite{Toni2009} and adaptive sequential Monte Carlo methods \cite{DelMoral2012}.

In this paper, we focus on the problem of summary statistics. In early works of ABC, summary statistics are chosen by domain experts in an ad-hoc manner. It is manageable if the dimensionality is small; which usually means that the model is well understood by the experts. But choosing a set of appropriate summary statistics is much more difficult in complex models. To address this problem, a set of redundant and hopefully sufficient summary statistics are often constructed in the first place, as initial summary statistics; dimension reduction methods are then applied yielding a set of low dimensional summary statistics while persevering the sufficiency or information. 

Many dimension reduction methods have been proposed for ABC. Entropy based subset selection\cite{Joyce2008}, partial least square\cite{Wegmann2009}, neural network\cite{Blum2009} and expected posterior mean\cite{Fearnhead2012} are a few of them. The entropy based subset selection method works well in instances where the set of low dimensional summary statistics is a subset of the initial summary statistics, but the computational complexity increases exponentially with the size of the initial summary statistics. The partial least square and neural network methods aim to capture the nonlinear relationships of the original summary statistics. In both cases, a specific form of the regression function is assumed. A comprehensive review \cite{Blum2013} discusses the methods mentioned above and compares the performances. While the results are a mixed bag, it is reported that the expected posterior mean method (Semi-automatic ABC) \cite{Fearnhead2012} produces relatively better results compared to the methods mentioned above in various experiments. It is a popular choice also due to its simplicity.

Semi-automatic ABC \cite{Fearnhead2012} uses the estimated posterior mean as summary statistics. A pilot run of ABC is first conducted to identify the regions of parameter space with non-negligible probability mass. The posterior mean is then estimated using the simulated data from that region and is used as the summary statistics in a following formal run of ABC. A linear model of the form: $\theta_i = {\beta^{(i)}}f(\mathbf{y}) + \epsilon_i$ is used in the estimation, where $f(y)$ are the possibly non-linear transforms of the data. For each application, the features $f(y)$ are carefully designed to achieve a good estimation. In practice, it may be difficult to determine a good set of features given a particular application. In these cases, a vector of powers of the data $\mathbf{(y,y^2,y^3,y^4,...)}$ is often used as noted in \cite{Fearnhead2012}.

To provide a principled way to design the regression function, capture the higher order non-linearity and realize an automatic construction of summary statistics, we introduce the kernel based sufficient dimension reduction method as an extension of the linear projection based Semi-automatic ABC. The dimension reduction used here is a localized version of Gradient based kernel dimension reduction (GKDR) \cite{Fukumizu2004}. GKDR, as a general dimension reduction method, estimates the projection matrix onto the sufficient low dimensional subspace by extracting the eigenvectors of the kernel derivatives matrices in the reproducing kernel Hilbert spaces (RKHS). We give a brief review of the method in Section 2. In addition to the GKDR, in which the estimation averages over all data points to reduce variance, a localized GKDR is proposed by averaging over a small neighborhood around the observation in ABC. Each point is weighted using a distance metric measuring the difference between the simulated data and the observation, similar to the distance kernel function in \eqref{eq:1}, to concentrate on the observation point. Another proposal is to use different summary statistics for different parameters. Note that sufficient subspace for different parameters can be different, depending on the particular problem. In these cases, as will be shown later by experiments, applying separated dimension reduction procedure yields a better estimation. 

The proposed method gives competitive results in comparison with Semi-automatic ABC\cite{Fearnhead2012} when using simple rejection sampling. Substantial improvements are reported in the sequential Monte Carlo cases, where threshold $\epsilon$ is pushed to as small as possible to isolate the performance of summary statistics from the Monte Carlo error.

The paper is organized as follows. In Section 2, we review GKDR and introduce its localized modification followed by discussions of computation considerations. In Section 3, we show simulation results for various commonly conducted ABC experiments, and compare the proposed method with the Semi-automatic ABC. 

\section{Local Kernel Dimension Reduction}
In this section, we review the Gradient based Kernel Dimension Reduction (GKDR) and propose the modified Local GKDR (LGKDR). Discussions are given at the end of this section.

\subsection{Gradient based kernel Dimension Reduction}
Given observation $(s,\theta)$, where $s \in \mathbb{R} ^m$ are initial summary statistics and $\theta \in \mathbb{R}$ is the parameter to be estimated in a specific ABC application. Assuming that there is a $d$-dimensional subspace $U \subset \mathbb{R} ^d$, $d<m$ such that

\begin{equation}\label{eq02}
\theta \independent s \mid B ^T s,
\end{equation}
where $B=(\beta_1,...,\beta_d)$ is the orthogonal projection matrix from $\mathbb{R} ^m$ to $\mathbb{R}^d$ . The columns of $B$ spans $U$ and $B^TB=\mathbf{I}_d$. Condition \eqref{eq02} shows that given $B^T s$, $\theta$ is independent of the initial summary statistics $s$. It is then sufficient to use $d$ dimensional constructed vector $z=B^Ts$ as the summary statistics. This subspace $U$ is called {\it effective dimension reduction(EDR)} space \cite{Li1991} in classical dimension reduction literatures. While there are a tremendous amount of published works about estimating the EDR space, in this paper, we propose to use GKDR in which no strong assumption of marginal distribution or variable type is made. The following is a brief review of GKDR, and for further details, we refer to \cite{Fukumizu2004} \cite{Fukumizu2009} \cite{Fukumizu2014}. 

Let $B=(\beta_1,...,\beta_d) \in \mathbb{R} ^{m \times d}$ be the projection matrix to be estimated, and $z=B^T s$. We assume \eqref{eq02} is true and $p(\theta|s)=\tilde{p}(\theta|z)$. The gradient of the regression function is denoted by $\nabla_s$ as 

\begin{equation}\label{eq03}
\nabla_s=\frac{\partial E(\theta|s)}{\partial s}=\frac{\partial E(\theta|z)}{\partial s}=B\frac{\partial E(\theta|z)}{\partial z}
\end{equation}
which shows that the gradients are contained in the EDR space. Given the following estimator  $M=E[\nabla_s\nabla_s^T]=BAB^T$, where $A_{ij}=E[E(\theta|\beta_i^Ts)E(\theta|\beta_j^Ts)]$, $i,j = 1,..., d$. The projection directions $\beta$ lie in the subspace spanned by the eigenvectors of $M$. It is then possible to estimate the projection directions using eigenvalue decomposition. In GKDR, the matrix $M$ is estimated by the kernel method described below.

Let $\Omega$ be an non-empty set, a real valued kernel $k:\Omega \times \Omega \rightarrow \mathbb{R}$ is called positive definite if $\sum_{i,j=1}^n c_i c_j k(x_i.x_j) \ge 0$ for any $x_i \in \Omega$ and $c_i \in \mathbb{R}$. Given a positive definite kernel $k$, there exists a unique reproducing kernel Hilbert space (RKHS) $H$ associated with it such that: (1)${k(\cdot, x)}$ spans $H$;(2)$H$ has the \emph{reproducing property}\cite{Aronszajn1950}: for all $x \in \Omega$ and $f\in H$, $\langle f, k(\cdot , x) \rangle = f(x)$.  

Given training sample $(s_1,\theta_1),...,(s_n,\theta_n)$, let $k_S(s_i,s_j)=exp(-||s_i-s_j||^2/\sigma_S^2)$ and $k_{\Theta}(\theta_i,\theta_j)=exp(-||\theta_i-\theta_j||^2/\sigma_{\Theta}^2)$ be Gaussian kernels defined on $\mathbb{R}^m$ and $\mathbb{R}$, associated with RKHS $H_S$ and $H_{\Theta}$, respectively. With assumptions of boundedness of the conditional expectation $E(\theta|S = s)$ and the average gradient functional with respect to $z$, the functional can be estimated using cross-covariance operators defined in RKHS and the consistency of their empirical estimators are guaranteed \cite{Fukumizu2014}. Using these estimators, we construct a covariance matrix of average gradients as 

\begin{equation}\label{eq04}
\widehat{M}_n(s_i) = \nabla \mathbf{k}_S(s_i)^T (G_S + n\epsilon_nI_n)^{-1}G_{\Theta}(G_S + n\epsilon_nI_n)^{-1}\nabla \mathbf{k}_S(s_i)
\end{equation}
where $G_S$ and $G_{\Theta}$ are Gram matrices $k_S(s_i,s_j)$ and $k_{\Theta}(\theta_i,\theta_j)$, respectively. $\nabla \mathbf{k}_S \in \mathbb{R}^{n\times m}$ is the derivative of the kernel $\mathbf{k}_S(\cdot,s_i)$ with respect to $s_i$, and $\epsilon_n$ is a regularization coefficient. This matrix can be viewed as the straight forward extension of covariance matrix in principle component analysis (PCA); the data here are the features in RKHS representing the gradients instead of the  gradients in their original real space.

The averaged estimator $\tilde{M} = 1/n \sum_{i=1}^n \widehat{M_n}(s_i)$ is calculated over the training sample $(s_1,\theta_1),...,(s_n,\theta_n)$. Finally, the projection matrix $B$ is estimated by taking $d$ eigenvectors corresponding to the $d$ largest eigenvalues of $\tilde{M}$ just like in PCA, where $d$ is the dimension of the estimated subspace.

\subsection{Local Modifications}
As discussed above, the estimator $\tilde{M}$ is obtained by averaging over the training sample $s_i$. When applied to ABC, since only one observation sample is available, we propose to generate a set of training data using the generating model and introduce a weighting mechanism to concentrate on the local region around the observation and avoid regions with low probability density.

Given simulated data $X_1,...,X_N$ and a weight kernel $K_w:\mathbb{R}^m \rightarrow \mathbb{R}$, we propose the local GKDR estimator

\begin{equation}\label{eq05}
\tilde{M}=\frac{1}{N} \sum_{i=1}^{N} K_w(X_i){\widehat{M}}(X_i)
\end{equation}
where $\widehat{M}$ is $m\times m$ matrix and $K_w(X_i)$ is the corresponding weight. $K_w(x)$ can be any weighting kernel. In the numerical experiments, a triweight kernel is used, which is written as

\begin{equation*}
K_w(X_i)=(1-u^2)^3\bm{1}_{u<1} \quad u=\frac{\|X_i-X_{obs}\|^2}{\|X_{th}-X_{obs}\|^2}
\end{equation*}
where $\bm {1}_{u<1}$ is the indicator function, and $X_{th}$ is the threshold value which determines the bandwidth. The normalization term of the triweight kernel is omitted since it does not change the eigenvectors we are estimating. The bandwidth determined by $X_{th}$ is chosen by empirical experiments and will be described in \ref{sec:hyper}. The Triweight kernel is chosen for its concentration in the central area than other "bell shaped" kernels and works well in our experiments. Other distance metrics could be used instead of squared distance.

Description of LGKDR algorithm are given in Algorithms~\ref{algo_gkdr}. Procedure \textbf{GenerateSample} is the algorithm to generate sample with parameter as input. Procedure \textbf{LGKDR} is the algorithm to calculate matrix $M(X_i)$ as given in (\ref{eq04}) and (\ref{eq05}). 

Since the dimension reduction procedure is done before the sampling, it works as a pre-processing unit to the main ABC sampling procedure. It can be embodied in any ABC algorithm using different sampling algorithms. In this paper, the rejection sampling method is firstly employed for its simplicity and low computation complexity as a baseline. Further results on Sequential Monte Carlo ABC are also reported to illustrate the advantage of the purposed method. In these experiments, the distance thresholds are pushed to as small as possible to suppress the Monte Carlo errors and isolate the effects of summary statistics alone. 

\begin{algorithm}
  \caption{LGKDR}\label{algo_gkdr}
  \SetKwFunction{GenerateSample}{GenerateSample}\SetKwFunction{LGKDR}{LGKDR}
  \SetKwInOut{Input}{input}\SetKwInOut{Output}{output}
  \Input{weighting kernel $K_w$, procedure \GenerateSample, prior distribution $D_{prior}$, number of accepted sample N, process \LGKDR}
  \Output{projection matrix $B$}
  \BlankLine
  \emph{training sample generation}\;
  \While{$i \le N$}{
    draw $\theta_i \leftarrow D_{prior}$\;
    $X_i \leftarrow \GenerateSample(\theta_i)$\;
    $w(i) \leftarrow K_w(X_i)$\;
      \If{$w \le 1$}{$i \leftarrow i + 1$}
    }
    \BlankLine
  \emph{calculate B}\;
  \For{$j \leftarrow 1$ \KwTo $N$}{$M \leftarrow M + \LGKDR(w(j).*X_j)$
    }
    $M_{ave} \leftarrow M./N$\;
    $B \leftarrow eigen(M_{ave})$\;
\end{algorithm}

\begin{algorithm}
  \caption{Rejection-ABC}\label{RABC}
  \SetKwInOut{Input}{input}\SetKwInOut{Output}{output}
  \Input{projection matrix $B$, distance kernel $K_d$, bandwidth $\epsilon$, number of sample $N_{ABC}$ and observation $X _{ob}$}
  \Output{set of parameters $\{\theta(j)\}$}
  \BlankLine
  $j \leftarrow 1$\;
  \For{$i \leftarrow 1$ to $N_{ABC}$}{
    draw $\theta_i \leftarrow D_{prior}$\;
    $X_i \leftarrow \GenerateSample(\theta_i)$\;
    \If{$K_d(B^TX_i,B^TX_{ob}) < \epsilon$}{$\theta(j) \leftarrow \theta_i$\;
    $j \leftarrow j +1$\;}
    }
\end{algorithm}

\begin{algorithm}
  \caption{Sequential-ABC}\label{SMC}
  \SetKwFunction{Resample}{Resample}\SetKwFunction{MoveParticle}{MoveParticle}
  \SetKwInOut{Input}{input}\SetKwInOut{Output}{output}  
  \Input{projection matrix $B$, distance kernel $K_d$, target threshold $\epsilon _t$, number of particle $N_{ABC}$, effective sample size threshold $ess_t$}
  \Output{set of parameters $\{\theta(j)\}$}
  \BlankLine 
  \For{$i \leftarrow 1$ to $N_{ABC}$}{
  	draw $\theta_i \leftarrow D_{prior}$\;
	$X_i \leftarrow \GenerateSample(\theta_i)$\;
  }
  $\epsilon \leftarrow Maximum(K_d(B^TX,B^TX_{obs}))$\;
  \While{$\epsilon \leq \epsilon _t$}{
  		decrease $\epsilon$\;
        \For{$i \leftarrow 1$ to $N_{abc}$}{
        	\If{$K_d(B^TX_i,B^TX_{abc}) \leq \epsilon$}{
            	$X_{partical} \leftarrow X_i$\;
                $\theta _{particle} \leftarrow \theta _i$\;
                calculate weight $W_i$\;
                } 
        }
        \MoveParticle($X_{particle}$)\;
        \If{$\sum _{i=0} ^{N_{abc}}W_i \leq ess_t$}{
        	$X \leftarrow $\Resample($X_{particle}$)\;}  
	}
  \BlankLine 
  \emph{MoveParticle}\;
  \For{$X_j\; in \; X_{particle}$}{
  	$\theta _{new} \leftarrow Normal(\theta_j,std(\theta _{partical}))$\;
    $X_j \leftarrow \GenerateSample(\theta_{new})$\;
    update weight $W_j$\;
  }
  \BlankLine 
  \emph{Resample}
  \For{$i \leftarrow 1$ to $N_{abc}$}{
  	copy $X_i$ $N_{abc}W_i$ times\;
    \If{$W_i = 0$}{
    	discard $X_i$}
    re-weight $W_i$ to $1/N_{abc}$}
\end{algorithm}

\subsection{Separated Dimension Reduction}
It is expected that separated construction of summary statistics for each parameter is beneficial for applications. Different information may be crucial for different parameters. If we estimate the projection directions separately for each parameter, the accuracy of the projection can be improved, and a lower dimensionality may be achievable. 

The LGKDR incorporates information of $\theta$ in the calculation of gradient matrix $\tilde{M}$. If $\theta$ is a vector, the relation of different elements of $\theta$ are contained in the gram matrix $G_{\theta}$ as in (\ref{eq04}). Separate estimations concentrate on the information of the specific parameter rather than the whole vector. As shown in the experiments in Section 3.2, it can construct significantly more informative summary statistics in some problems by means of reducing estimation error.

For Semi-automatic ABC\cite{Fearnhead2012}, the summary statistics for each parameter is the estimated posterior mean, thus naturally separated. Using separated summary statistics means using a 1 dimensional summary statistics for each parameter. our experiments do not show good results using this setting. For best subset selection methods\cite{Wegmann2009}\cite{Nunes2010}, summary statistics are chosen as the best subset of the original summary statistics using mutual information or sufficiency criterion. It can also be extended to a separated selection procedure. In LGKDR, we simply construct different summary statistic by using only the particular parameter as the response variable.

\subsection{Discussion on hyper parameters}
\label{sec:hyper}
Several parameters need to be set for achieving good results in LGKDR ABC. Parameters for the sampling procedures will be discussed in the experiments section. In this section, the parameters of the LGKDR part is explained.

First, the bandwidth of the weighting kernel, which measures the degree of concentration of weightings of projection directions, affects the accuracy of LGKDR. By selecting a large bandwidth, the weights of directions spread out a larger region around the observation points. A small bandwidth concentrates the weights on the directions estimated close to the observation sample. In our experiments, a bandwidth corresponding to an acceptance rate of approximately $10\%$ gives a good result and is used throughout the experiments. The same parameter is set for the Semi-automatic ABC as well for the similar purpose. A more principled method for choosing bandwidth, like cross validation, could be applied to select the acceptance rate if the corresponding computation complexity is affordable. 

The bandwidth parameters of the Gaussian kernels $\sigma_S$, $\sigma_{\Theta}$ and the regularization parameter $\epsilon_n$ are the configuring parameters that are crucial to all kernel based methods. The first two determine the function spaces associated with the positive definite kernels and the regularization parameter affects the convergence rate (see \cite{Smola1998}). In our method, cross validation is adopted to select the proper parameters for different experiments. In the cross validation, for each set of candidate parameters, the summary statistics are constructed using a simulated observation $\theta_{obs}, s_{obs}$, a training set $(\theta_{training},S_{training})$ and a test set $(\theta_{test},S_{test})$. A small pilot run of rejection ABC is performed and the estimation of parameters are calculated by kNN regression of $\theta_{test}$ with the $S_{test}$. $\it{K}$ is set to 5 in all cases. The set of parameters that yield the smallest least error between the $\theta_{test}$ and $\theta_{obs}$ are chosen. The final summary statistics are constructed using these chosen parameters and are passed to the formal run of ABC subsequent to the summary construction procedure.

\subsection{Computational Complexity}
Computational complexity is one central concern of ABC methods. LGKDR is more computationally demanding than the linear regression-type methods. It requires matrix inversion and solving eigenvalue problems and above all, the cross validation procedure. The actual complexity depends on the training sample size used. For the experiments shown in this paper, the training sample size are fixed to $2\times10^3$ and $10^4$ for LGKDR and Semi-automatic ABC, respectively. Under this setting, the total computational time of LGKDR are about 10 times over the linear regression. While the computational complexity is higher, it is a necessary price to pay if the non-linearity between the initial summary statistics is strong. For these cases, being unable to capture the non-linear information in summary statistics would induce a poor sampling performance, which leads to a biased estimation. Meanwhile, if the generating model itself is complex, the computational time used on the LGKDR will become less significant. Finally, although the cross validation procedure takes the majority of computation time in LGKDR, it needs to be performed only once for each problem to fix the parameters. Once the parameters are chosen, the computation of LGKDR is comparable to the linear-type algorithms. Overall the computational complexity of an ABC method depends on both the summary statistics calculation step and the following sampling step. For complex models like ones in population genetics, sampling is significantly more time consuming than the dimension reduction procedure whichever we use.

\section{Experiments}\label{Experiments}
In this section, we investigate three problems to demonstrate the performance of LGKDR. Our method is compared to the classical ABC using initial summary statistics and the Semi-automatic ABC \cite{Fearnhead2012} using estimated posterior means. In the first problem, we discuss a population genetics model, which was investigated in many ABC literatures. We adopt the initial summary statistics used in \cite{Nakagome2013}, and rejection ABC is used as the sampling algorithm. In the second problem, a M/G/1 stochastic queue model which was used in \cite{Blum2009} and \cite{Fearnhead2012} are discussed. While the model is very simple, the likelihood function could not be trivially computed. In the last experiment we explore the Ricker model as discussed in \cite{Wood2010} and \cite{Fearnhead2012}. The latter two problems are investigated by both Rejection ABC and sequential Monte Carlo ABC method (SMC ABC) \cite{DelMoral2012}, the first problem is omitted from SMC ABC because it involves repeated calling an outside program for simulation and is too time consuming for SMC ABC. 

\subsection{Implementation Details}
The Rejection ABC is described in Algorithm \ref{RABC} and the SMC ABC is shown in Algorithm \ref{SMC}. The hyper-parameters used in LGKDR is set as discussed in section \ref{sec:hyper}. We use a modified code from \cite{DelMoral2012} and R package "Easyabc" \cite{code1} in our SMC implementation and would like to thank the corresponding authors. Gaussian kernels are used in all the LGKDR algorithms. The detailed specifications of Semi-automatic ABC will be described in each experiment.

For evaluation of the experiments conducted using rejection ABC, a set of parameters $\theta^j$ where $j\in 1,...,N_{obs}$ and corresponding observation sample $Y_{obs}^j$ are simulated from the prior and the conditional probability $p(Y|\theta)$, respectively, and used as the observations. For each experiment, we fix the total number of simulations $N$, and the number of accepted sample $N_{acc}$. The sample used for rejection are then generated and fixed for all three methods. In this case we can ignore the randomness in the simulation program and accurately determine the acceptance rate, which is the single most influential parameter for estimation accuracy. The Mean squared error (MSE) over the accepted parameters $\hat{\theta}_i^j$ and observation $\theta^j$ are defined as

\begin{equation*}
MSE_j = \frac{1}{N_{acc}} \left (\sum_{i=1}^{N_{acc}} (\theta^j -\hat{\theta}_i^j)^2\right) 
\end{equation*}

The Averaged Mean Square Error (AMSE) is then computed as the average over $MSE_j$ of each observation pair $(\theta^j, Y_{obs}^j)$ as

\begin{equation*}
AMSE = \frac{1}{N_{obs}} \sum _{j=1}^{N_{obs}} MSE_j.
\end{equation*}
It is used as the benchmark for Rejection ABC. Because of the difference of computation complexity, for fairness of comparison, the acceptance rates are set differently. For LGKDR, the acceptance rate is set to $1\%$; while for Semi-automatic ABC and original ABC, the acceptance rates are set to $0.1\%$. The training sample and simulated sample are generated from the same prior and remain fixed.

For SMC ABC, to get to as small tolerance as possible, the simulation time is different for different method. AMSE is used as benchmark for accuracy and computational time are reported for each experiment. In the case of Ricker model, due to the extremely long simulation time, only one observation is used and AMSE is just MSE in this case. 

\subsection{Parameter Settings}
Several parameters are necessary in running the simulations in ABC. For Rejection ABC, the total number of samples $N$ and the accepted number of samples $N_{acc}$ are set before the simulation as mentioned above. For Semi-automatic ABC and LGKDR, a training set needs to be simulated to calculate the projection matrix. For LGKDR, a further testing set is also generated for cross validation purposes. The value of these parameters are reported in the corresponding experiments. The simulation time for generating these sample set are negligible compared to the main ABC, especially in SMC ABC. For LGKDR, another important parameter is the target dimensionality $D$. There is no theoretically sound methods available to determine the intrinsic dimensionality of the initial summary statistics; yet in practice, since the projection matrix consists of extracted eigenvectors of the matrix $M$ as in (\ref{eq05}), it is straightforward to test the performance of LGKDR using different projection matrices $B$ with different setting dimensionality and determine a proper dimensionality directly. In our experiments, we run several rejection ABC using different $B$, and fix the dimensionality. A starting point can be set by preserving $70\%$ of the largest eigenvalues in magnitude and it usually works well. There are a large collection of literatures on how to choose the number of principle components in PCA, which is similar to our problem, for example, see \cite{sergio1999} and reference therein.

\subsection{Population Genetics}
Analysis of population genetics is often based on the coalescent model\cite{Hudson2002b}. A constant population model is used in simple situations, where the population is assumed unchanged across generations. The parameter of interests in this case is the scaled mutation rate $\theta$, which controls the probability of mutation between each generation. The detailed introduction of coalescent models can be found in \cite{Nordborg2008}. Various studies \cite{Beaumont2009} \cite{Marjoram2003} \cite{Sisson2007} have been conducted in population genetics following different sampling algorithms. In this study, we adopt the setting of kernel ABC \cite{Nakagome2013} and compare the performance with ABC and Semi-automatic ABC. 

100 chromosomes are sampled from a constant population $(N=10000)$. The summary statistics are defined using the spectrum of the numbers of segregating sites, $\textbf{s}_{sfs}$, which is a coarse-grained spectrum consisting of 7 bins based on the Sturges formula $(1 + log_2S_{seg})$. The frequencies were binned as follows: $0 - 8\%, 8 - 16\%, 16 - 24\%, 24 - 32\%, 32 - 40\%, 40 - 48\%$ and $48 - 100\%$, we use the uniform distribution $\theta \sim [0,30]$ in this study rather than the log-normal distribution in \cite{Nakagome2013}. As ABC is often used for exploratory researches, we believe that the performance based on an uninformative prior is important for evaluating summary statistics. The program package ms is used to generate the sample, which is of common choice in literature of coalescent model \cite{Hudson2002a}. 

We test 3 typical scaled mutation rates $5,8$ and $10$ rather than random draws from the prior. The results are averaged over 3 tests. A total number of $10^6$ sample is generated; $10^5$ sample is generated as the training sample for LGKDR and Semi-automatic ABC. Different acceptance rates are set for different methods as discussed above. We use $\textbf{s}_{sfs}$ as the summary statistics for both Semi-automatic ABC and LGKDR. Local linear regression is used as the regression function for the former. In LGKDR, the dimension is set to 2.

As shown in Table-\ref{tab:coal}, the performance of both LGKDR and Semi-automatic ABC improve over original ABC method. LGKDR and Semi-automatic ABC achieve very similar results suggesting that the linear construction of summary statistics are sufficient for this particular experiment.

\begin{table}
\caption{Coalescent Model.}
\begin{center}
{
\begin{tabular}[l]{@{}lr}
  \hline
  Method & mutation rate $\theta$ \\ 
  ABC & 1.94 \\
  Semi-automatic ABC & 1.62 \\
  LGKDR & 1.66 \\
  \hline
  \end{tabular}
}
\end{center}
\label{tab:coal}
 \end{table}

\subsection{M/G/1 Queue Model}
The M/G/1 model is a stochastic queuing model that follows the first-come-first-serve principle. The arrival of customers follows a Poisson process with intensity parameter $\lambda$. The service time for each customer follows an arbitrary distribution with fixed mean (G), and there is a single server (1). This model has an intractable likelihood function because of its iterative nature. However a simulation model with parameter $(\lambda, \mu)$ can be easily implemented to simulate the model. It has been analyzed by ABC using various different dimension reduction methods as in \cite{Fearnhead2012} and \cite{Blum2009}, with comparison to the indirect inference method. We only compare our method with Semi-automatic ABC, since it produces substantially better results then the other methods mentioned above.

The generative model of the M/G/1 model is specified by

\begin{equation*}
Y_n = 
\begin{cases}
U_n & \text{if } \sum _{i=1} ^{n} W_i \le \sum _{i=1} ^{n-1}Y_i\\
U_n + \sum _{i=1} ^n W_i - \sum_{i=1} ^{n-1} Y_i & \text{if } \sum _{i=1} ^{n} W_i > \sum _{i=1} ^{n-1} Y_i
\end{cases}
\end{equation*}
where $Y_n$ is the inter-departure time, $U_n$ is the service time for the $n$th customer, and $W_i$ is the inter-arrival time. The service time is uniformly distributed in interval $[\theta_1,\theta_2]$. The inter-arrival time follows an exponential distribution with rate $\theta_3$. These configurations stay the same as \cite{Blum2009} and \cite{Fearnhead2012}. We set uninformative uniform priors for $\theta_1, \theta_2-\theta_1$ and $\theta_3$ as $[1,10]^2\times[1,1/3]$.

For the rejection ABC, we simulate a set of 30 pairs of $(\theta_1,\theta_2,\theta_3)$ but avoid boundary values. They are used as the true parameters to be estimated. The total number of $10^6$ sample are generated. The posterior mean is estimated using the empirical mean of the accepted sample. The simulated sample are fixed across different methods for comparison. 

we use the quantiles of the sorted inter-departure time $Y_n$ as the exploration variable of the regression model $f(y)$ as in \cite{Fearnhead2012}. The powers of the variables are not included as no significant improvements are reported. A pilot ABC procedure is conducted using a fixed training sample set of size $10^4$. Local linear regression is used rather than a simple linear regression for better results. For LGKDR, we use the same quantiles as initial summary statistics for dimension reduction as in Semi-automatic ABC. The number of accepted training sample is $2\times 10^3$ in for the LGKDR. The dimension is manually set to 4, as small as the performance is not degraded.  
 
The experimental results of Rejection ABC are shown in Table-\ref{tab:queue}. ``LGKDR" refers to the LGKDR that does not use separated estimation. ``focus 1" denotes the separated dimension reduction for parameter $\theta_1$, and the following rows are of similar form. Compared to ABC, ``Semi-automatic ABC" gives substantial improvement on the estimation of $\theta_1$; the other parameters show similar or slightly worse results. LGKDR method improves over ABC on $\theta_1$ and $\theta_2$, but the estimation of $\theta_1$ is not as good as in Semi-automatic ABC. However, after applying separated estimation, $\theta_1$ presents a substantial improvement compared to Semi-automatic ABC. Separated estimations for $\theta_2$ and $\theta_3$ give no improvements. It suggests that the sufficient dimension reduction subspace for $\theta_1$ is different from the others and a separated estimation of $\theta_1$ is necessary.

For SMC ABC, a set of 10 pairs of parameters are generated, and results on SMC and LGKDR are reported. all other setting are same as rejection ABC. We omit the results of using Semi-automatic ABC since the sequential chain did not converge properly using these summary statistics and the induced error was too large to be meaningful. In SMC ABC, two experiments are reported: SMC ABC1 and SMC ABC2. The number of particles are set to $2\times 10^4$ and $10^5$, respectively. In LGKDR, the number of particles are set to $2\times 10^4$ and the training sample size for the calculation of projection matrix is $2\times 10^3$, accepted from a training set of size $4\times 10^4$ simulated samples using rejection ABC. The dimensionality is set to 5. Cross validation is conducted using a test set of size $2\times 10^4$. 

Results of SMC ABC are shown in Table-\ref{tab:queue1}. AMSEs are reported. The simulation time is shown as well. The computational time of constructing LGKDR summary statistics is included in the total simulation time and listed in the bracket. The results show that LGKDR gives better results of parameter $\theta _1$ and $\theta _2$, using less time compared to SMC ABC with set $E2$. The estimation of $\theta _3$ is worse but the difference is small ($0.005$). Focusing on $\theta _3$ produces an estimation as good as in SMC ABC.

\begin{table}
\caption{M/G/1 Queue Model, Rejection ABC} \label{tab:queue} 
\begin{center}
{
\begin{tabular}[l]{@{}lrrr}
  \hline
  Method & $\theta_1$ & $\theta_2$ & $\theta_3$ \\
   ABC & 0.2584 & 0.5113 & 0.0019 \\
   Semi-automatic ABC & 0.0112 & 0.5279 & 0.0024 \\
   LGKDR & 0.0623 & 0.2259 & 0.0023 \\
   LGKDR(focus 1) & 0.0082 & 5.0656 & 0.0031 \\
   LGKDR(focus 2) & 0.3942 & 0.2514 & 0.0020 \\
   LGKDR(focus 3) & 0.2229 & 3.4958 & 0.0020 \\
 \end{tabular}
}
 \end{center}
 \end{table}

\begin{table}
\caption{M/G/1 Queue Model, SMC ABC} \label{tab:queue1} 
\begin{center}
{
\begin{tabular}[l]{@{}lrrrr}
  \hline
   Method & $\theta_1$ & $\theta_2$ & $\theta_3$ & Total time\\
   SMC ABC 1 & 0.0404 & 0.4928 & 0.0139 & 9.6e+03 \\
   SMC ABC 2 & 0.0429 & 0.1964 & 0.0054 & 3.3e+04 \\  
   LGKDR & 0.0235 & 0.1605 & 0.0110 & 2.0e+04 (7.78e+3)\\   
   LGKDR(focus 2) & 0.4854 & 0.1383 & 0.0059 & 2.1e+04 (7.85e+3)\\
   \hline
 \end{tabular}
}
 \end{center}
 \end{table}

\subsection{Ricker Model}
Chaotic ecological dynamical systems are difficult for inference due to its dynamic nature and the noises presented in both the observations and the process. Wood \cite{Wood2010} addresses this problem using a synthetic likelihood inference method. Fearnhead \cite{Fearnhead2012} tackles the same problem with a similar setting using the Semi-automatic ABC and reports a substantial improvement over other methods. In this experiment, we adopt the same setting and apply LGKDR with various configurations.

A prototypic ecological model with Richer map is used as the generating model in this experiment. A time course of a population $N_t$ is described by
\begin{equation}\label{eq:ricker}
N_{t+1}=rN_te^{-N_{t}+e_t}
\end{equation}
where $e_t$ is the independent noise term with variance $\sigma_e^2$, and $r$ is the growth rate parameter controlling the model dynamics. A Poisson observation $y$ is made with mean $\phi N_t$. The parameters to infer are $\theta = (log(r), \sigma_e^2,\phi)$. The initial state is $N_0 = 1$ and observations are $y_{51} , y_{52} , \cdots , y_{100}$. 

The original summary statistics used by Wood\cite{Wood2010} are the observation mean $\bar{y}$, auto-covariances up to lag 5, coefficients of a cubic regression of the ordered difference $y_t - y_{t-1}$ on the observation sample, estimated coefficients for the model $y_{t+1}^{0.3} = \beta_1 y_t^{0.3} + \beta_2 y_t^{0,6} + \epsilon_t$ and the number of zero observations $\sum_{t=51}^{100} \bm {1}(y_t=0)$. This set is denoted as E0 as in \cite{Fearnhead2012}. Additional two sets of summary statistics are defined for Semi-automatic ABC. The smaller E1 contains E0 and $\sum_{t=51}^{100} \bm{1}(y_t=j)$ for $1\le j \le 4$, logarithm of sample variance, $log(\sum_{t=51}^{100}y_t^j)$ for $2 \le j \le 6$ and auto-correlation to lag 5. Set E2 further includes time-ordered observation $y_t$, magnitude-ordered observation $y_{(t)}$, $y_t^2$, $y_{(t)}^2$, $\{log(1+y_t)\}$, $\{log(1+y_{(t)})\}$, time difference $\Delta y_t$ and magnitude difference $\Delta y_{(t)}$. Additional statistics are added to explicitly explore the non-linear relationships of the original summary statistics and are carefully designed. 

In Rejection ABC, we use set E0 for ABC without dimension reduction since the dimension of the larger sets induces severely decreased performance. Sets E1 and E2 are used for Semi-automatic ABC as in \cite{Fearnhead2012}. In LGKDR, we tested sets E0 and E1 in different experiments. The result on E2 is omitted as the result is similar with using the smaller set of statistics, indicating that manually designed non-linear features are unnecessary for LGKDR. The sufficient dimension is set to $5$; a smaller value induces substantial worse results. We simulated a set of $30$ parameters, a fixed simulated sample of size $10^7$ for all the methods and a training sample of size $10^6$, a test sample of size $10^5$ for LGKDR and Semi-automatic ABC. The values of $log(r)$ and $\phi$ are fixed as in \cite{Fearnhead2012}, and $log(\sigma_e)$ are drawn from an uninformative uniform distribution on $[log(0.1), 0]$. 

The results are shown in Table \ref{tab:ricker}. The performance of Semi-automatic ABC using the bigger set E2 is similar to ABC but is substantially worsen with set E1, suggesting that the non-linear information are essential for an accurate estimation in this model. These features are needed to be explicitly designed and incorporated into the regression function for Semi-automatic ABC. LGKDR using summary statistics set E0 gives similar results compared with ABC. Using larger set E1, the accuracy of $log(r)$ is slightly worse than using set E0, but the accuracy of $\sigma_e$ and $\phi$ present substantial improvements. The additional gains of separate constructions of summary statistics in this model are mixed for different parameter, $log(r)$ and $\phi$ show very small improvements but $\sigma_e$ gets improvements in both cases. Overall, We recommend using separate constructions for the potential improvements if the additional computational costs are affordable. 

In SMC ABC, we use set E0 for the SMC, E1 for LGKDR and both E1 and E2 for Semi-automatic ABC. Number of particles is set to $5\times 10^3$ for all experiments. Other parameters are the same as in Rejection ABC. Only one set of parameter is used and the time of simulation is set to achieve as small as possible tolerance. Simulation time is reported with computational time of LGKDR included. We also test several settings of dimensionality in LGKDR to check the influence of that hyper-parameter. As can be observed in the results, if the dimensionality is set too high, the efficiency of the SMC chain is decreased; if it is set too low, more bias is induced in the estimated posterior mean suggesting loss of information in the constructed summary statistics. In this experiment, dimensionality 6 is chosen by counting the number of largest $70\%$ eigenvalues in magnitude as discussed before. 

The results are shown in Table-\ref{tab:ricker1}, which shows that, if a proper dimensionality is chosen, LGKDR can achieve the similar results as Semi-automatic ABC using only $1/10$ simulation time.  

\begin{table}
\caption{Ricker Model, Rejection ABC}
\begin{center}
{
\begin{tabular}[l]{@{}lccc}
\hline
Method & $log(r)$ & $\sigma_e$ & $\phi$ \\ 
ABC(E0) & 0.049 & 0.217 & 0.944 \\
Semi-automatic ABC(E2) & 0.056 & 0.246 & 0.936 \\
Semi-automatic ABC(E1) & 0.082 & 0.279 & 1.387 \\
LGKDR(E0) & 0.043 & 0.241 & 0.984 \\
LGKDR(E0,focus1) & 0.043 & 0.221 & 1.221\\
LGKDR(E0,focus2) & 0.068 & 0.200 & 1.234\\
LGKDR(E0,focus3) & 0.047 & 0.211 & 1.007\\
LGKDR(E1) & 0.047 & 0.179 & 0.895 \\
LGKDR(E1,focus1) & 0.048 & 0.220 & 1.38\\
LGKDR(E1,focus2) & 0.059 & 0.174 & 2.694\\
LGKDR(E1,focus3) & 0.054 & 0.292 & 0.829\\
\hline
\end{tabular}
}
\end{center}
\label{tab:ricker}
\end{table}

\begin{table}
\caption{Ricker Model, SMC ABC}
\begin{center}
{
\begin{tabular}[l]{@{}lcccc}
\hline
Method & $log(r)$ & $\sigma_e$ & $\phi$ & Total time \\ 
ABC(E0) & 0.001 & 0.003 & 0.430 & 4.0e+5\\
Semi-automatic ABC(E2) & 0.002 & 0.020 & 0.013 & 4.3e+5 \\
Semi-automatic ABC(E1) & 0.031 & 0.079 & 0.019 & 1.7e+5 \\
LGKDR(Dimensional 3) & 0.024 & 0.131 & 0.779 & 8.6e+4\\
LGKDR(Dimensional 6) & 0.006 & 0.018 & 0.012 & 4.5e+4 \\
LGKDR(Dimensional 9) & 0.001 & 0.040 & 0.250 & 2.8e+5 \\
\hline
\end{tabular}
}
\end{center}
\label{tab:ricker1}
\end{table}

\section{Conclusions}\label{conclusions}
We proposed the LGKDR algorithm for automatically constructing summary statistics in ABC. The proposed method assumes no explicit functional forms of the regression functions nor the marginal distributions, and implicitly incorporates higher order moments up to infinity. As long as the initial summary statistics are sufficient, our method can guarantee to find a sufficient subspace with low dimensionality. While the involved computation is more expensive than the simple linear regression used in Semi-automatic ABC, the dimension reduction is conducted as the pre-processing step and the cost may not be dominant in comparison with a computationally demanding sampling procedure during ABC. Another advantage of LGKDR is the avoidance of manually designed features; only initial summary statistics are required. With the parameter selected by the cross validation, construction of low dimensional summary statistics can be performed as in a black box. For complex models in which the initial summary statistics are hard to identify, LGKDR can be applied directly to the raw data and identify the sufficient subspace. We also confirm that construction of different summary statistics for different parameter improve the accuracy significantly.

\bibliographystyle{gSCS}
\bibliography{GkdrABC}

\begin{thebibliography}{10}
\providecommand{\url}[1]{\normalfont{#1}}
\providecommand{\urlprefix}{Available from: }

\bibitem{Pritchard1999}
Pritchard~JK, Seielstad~MT, Perez-Lezaun~A, Feldman~MW. {Population growth of
  human Y chromosomes: a study of Y chromosome microsatellites.} Molecular
  biology and evolution. 1999 Dec;\hspace{0pt}16(12):1791--1798.

\bibitem{Beaumont2002}
Beaumont~MA, Zhang~W, Balding~DJ. {Approximate Bayesian computation in
  population genetics.} Genetics. 2002 Dec;\hspace{0pt}162(4):2025--2035.

\bibitem{Toni2009}
Toni~T, Welch~D, Strelkowa~N, Ipsen~A, Stumpf~MP. {Approximate Bayesian
  computation scheme for parameter inference and model selection in dynamical
  systems}. Journal of the Royal Society Interface.
  2009;\hspace{0pt}6(31):187--202.

\bibitem{Csillery2010}
Csilléry~K, Blum~MGB, Gaggiotti~OE, François~O. {Approximate Bayesian
  computation (ABC) in practice}. Trends in Ecology and Evolution.
  2010;\hspace{0pt}25(7):410--418.

\bibitem{Grelaud2009}
Grelaud~A, Robert~CP, Marin~JM, Rodolphe~F, Taly~JF. {ABC likelihood-free
  methods for model choice in Gibbs random fields}. Bayesian Analysis. 2009
  Jun;\hspace{0pt}4(2):317--335.

\bibitem{Bertorelle2010}
Bertorelle~G, Benazzo~A, Mona~S. {ABC as a flexible framework to estimate
  demography over space and time: some cons, many pros.} Molecular ecology.
  2010 Jul;\hspace{0pt}19(13):2609--2625.

\bibitem{Blum2010}
Blum~MGB. {Approximate Bayesian Computation: A Nonparametric Perspective}.
  Journal of the American Statistical Association.
  2010;\hspace{0pt}105(491):1178--1187.

\bibitem{Moore1995}
Moore~WS. {Inferring Phylogenies from Mtdna Variation - Mitochondrial-Gene
  Trees Versus Nuclear-Gene Trees}. Evolution. 1995;\hspace{0pt}49(4):718--726.

\bibitem{Marjoram2003}
Marjoram~P, Molitor~J, Plagnol~V, Tavare~S. {Markov chain Monte Carlo without
  likelihoods}. Proc Natl Acad Sci US A.
  2003;\hspace{0pt}100(0027-8424):15324--15328.

\bibitem{Sisson2007}
Sisson~SA, Fan~Y, Tanaka~MM. {Sequential Monte Carlo without likelihoods.}
  Proceedings of the National Academy of Sciences of the United States of
  America. 2007;\hspace{0pt}104(6):1760--1765.

\bibitem{Joyce2008}
Joyce~P, Marjoram~P. {Approximately sufficient statistics and bayesian
  computation.} Statistical applications in genetics and molecular biology.
  2008;\hspace{0pt}7(1):Article26.

\bibitem{Wegmann2009}
Wegmann~D, Leuenberger~C, Excoffier~L. {Efficient approximate Bayesian
  computation coupled with Markov chain Monte Carlo without likelihood}.
  Genetics. 2009;\hspace{0pt}182(4):1207--1218.

\bibitem{Blum2009}
Blum~MGB, Fran{¥c{c}}ois~O. Non-linear regression models for approximate
  bayesian computation. Statistics and Computing.
  2010;\hspace{0pt}20(1):63--73;
  \urlprefix\url{http://dx.doi.org/10.1007/s11222-009-9116-0}.

\bibitem{Fearnhead2012}
Fearnhead~P, Prangle~D. {Constructing summary statistics for approximate
  Bayesian computation: Semi-automatic approximate Bayesian computation}.
  Journal of the Royal Statistical Society Series B: Statistical Methodology.
  2012;\hspace{0pt}74(3):419--474.

\bibitem{Blum2013}
Blum~MGB, Nunes~MA, Prangle~D, Sisson~SA. {A Comparative Review of Dimension
  Reduction Methods in Approximate Bayesian Computation}. Statistical Science.
  2013 May;\hspace{0pt}28(2):189--208.

\bibitem{Fukumizu2014}
Fukumizu~K, Leng~C. {Gradient-Based Kernel Dimension Reduction for Regression}.
  Journal of the American Statistical Association.
  2014;\hspace{0pt}109(505):359--370;
  \urlprefix\url{http://www.tandfonline.com/doi/abs/10.1080/01621459.2013.838167}.

\bibitem{Li1991}
Li~KC. {Sliced Inverse Regression for Dimension Reduction}. Journal of the
  American Statistical Association. 1991;\hspace{0pt}86(414):316--327.

\bibitem{Fukumizu2004}
Fukumizu~K, Bach~FR, Jordan~MI. {Dimensionality Reduction for Supervised
  Learning with Reproducing Kernel Hilbert Spaces}. Journal of Machine Learning
  Research. 2004;\hspace{0pt}5(1):73--99;
  \urlprefix\url{http://portal.acm.org/citation.cfm?id=1005332.1005335}.

\bibitem{Fukumizu2009}
Kenji~Fukumizu~MIJ~Francis R~Bach. Kernel dimension reduction in regression.
  The Annals of Statistics. 2009;\hspace{0pt}37(4):1871--1905;
  \urlprefix\url{http://www.jstor.org/stable/30243690}.

\bibitem{Aronszajn1950}
Aronszajn~N. {Theory of reproducing kernels}. Transactions of the American
  Mathematical Society. 1950;\hspace{0pt}68(3):337--337.

\bibitem{Nunes2010}
Nunes~M, Balding~DJ. {On optimal selection of summary statistics for
  approximate Bayesian computation.} Statistical applications in genetics and
  molecular biology. 2010;\hspace{0pt}9(1):Article34.

\bibitem{Nakagome2013}
Nakagome~S, Fukumizu~K, Mano~S. {Kernel approximate Bayesian computation in
  population genetic inferences}. Statistical Applications in Genetics and
  Molecular Biology. 2013;\hspace{0pt}12(6):667--678.

\bibitem{Wood2010}
Wood~SN. {Statistical inference for noisy nonlinear ecological dynamic
  systems.} Nature. 2010;\hspace{0pt}466(7310):1102--1104.

\bibitem{Hudson2002b}
Hudson~RR. {Generating samples under a Wright-Fisher neutral model of genetic
  variation.} Bioinformatics (Oxford, England).
  2002;\hspace{0pt}18(2):337--338.

\bibitem{Nordborg2008}
Nordborg~M. Coalescent theory. John Wiley \& Sons, Ltd; 2008. p. 843--877;
  \urlprefix\url{http://dx.doi.org/10.1002/9780470061619.ch25}.

\bibitem{Beaumont2009}
Beaumont~Ma, Cornuet~JM, Marin~JM, Robert~CP. {Adaptive approximate Bayesian
  computation}. Biometrika. 2009;\hspace{0pt}96(4):983--990;
  \urlprefix\url{http://biomet.oxfordjournals.org/cgi/doi/10.1093/biomet/asp052}.

\bibitem{Hudson2002a}
Hudson~R. {Ms a Program for Generating Samples Under Neutral Models}.
  Bioinformatics. 2002;\hspace{0pt}18(2002):337--338.

\end{thebibliography}

\end{document}